\documentclass[runningheads]{llncs}

\usepackage[T1]{fontenc}
\usepackage{amsmath}
\usepackage{graphicx}
\usepackage{verbatim}
\usepackage{subcaption}
\usepackage{algorithm}
\usepackage{algpseudocode}
\usepackage{microtype}
\usepackage{orcidlink}
\usepackage{amssymb}
\usepackage{float} 
\usepackage{tikz}
\usetikzlibrary{arrows.meta,positioning,fit,backgrounds,calc}
\begin{document}

\title{AML-QKD: Adaptive Machine Learning Framework for Real-time Parameter Tuning in QKD}
\titlerunning{AML-QKD: Adaptive ML Framework for Parameter Tuning in QKD}

\author{Noureldin Mohamed\inst{1}\orcidlink{0009-0001-4150-8690} \and
Jawaher Kaldari\inst{1}\orcidlink{0009-0000-5265-5373} \and
Saif Al-Kuwari\inst{1}\orcidlink{0000-0002-4402-7710}}

\authorrunning{}

\institute{Qatar Center for Quantum Computing, College of Science and Engineering, \\
Hamad Bin Khalifa University, Doha, Qatar \\
\email{nomo89098@hbku.edu.qa}}

\maketitle

\begin{abstract}
Despite the robust security guarantees of Quantum Key Distribution (QKD), practical deployment is hindered by dynamic channel noise and complex parameter optimization. We propose AML-QKD, a protocol-agnostic machine learning framework designed to maximize the Secure Key Rate (SKR) and minimize the Quantum Bit Error Rate (QBER) across the BB84, E91, and COW protocols. AML-QKD integrates Temporal Convolutional Networks (TCNs) for short-horizon forecasting of channel fluctuations, using a Proximal Policy Optimization (PPO) agent for real-time parameter tuning, while strictly adhering to composable security constraints. Simulations under realistic depolarizing and amplitude-damping noise demonstrate a 14---25\% increase in median SKR and a reduction in median QBER from 3.0\% to 1.5\%. Furthermore, an exploratory Quantum Reinforcement Learning (QRL) extension reveals a distinct quantum advantage for entanglement-based protocols (E91), achieving a 29.2\% throughput gain by natively processing non-local correlations. Our findings suggest that AML-QKD can offer a potentially resilient, security-preserving control architecture for next-generation quantum networks.

\end{abstract}

\section{Introduction}
Quantum Key Distribution (QKD) is at the forefront of secure communication technologies, offering unparalleled security by harnessing the principles of quantum mechanics. Unlike classical encryption methods, QKD provides security based on the fundamental laws of physics, making it immune to computational attacks. However, the practical deployment of QKD protocols faces significant challenges that stem from various operational and environmental factors. 

Among the most widely studied QKD protocols are BB84, E91, and COW. However, each protocol exhibits unique vulnerabilities in dynamic environments. For instance, BB84 relies on photon polarization \cite{bennett1984quantum}, where basis misalignment and detector dark counts drastically inflate the Quantum Bit Error Rate (QBER). The E91 protocol leverages quantum entanglement to ensure security \cite{ekert1991quantum}, but this entanglement is highly susceptible to channel noise, which degrades the CHSH inequality violation. Finally, the Coherent One-Way (COW) protocol \cite{Stucki:09} is constrained by phase drift between consecutive pulses, requiring precise phase-time locking to maintain interferometric visibility. Across all protocols, mitigating hardware- and channel-induced errors is critical to maximizing the secure key rate (SKR).

Traditional optimization approaches in QKD face several critical challenges \cite{scarani2009security_rmp}. First, protocol-specific parameter spaces dictate that each protocol requires a unique tuning strategy---from continuous polarization-axis adjustment in BB84 to density-matrix optimization in E91 to precise time-bin locking in COW. Second, noise-resilience trade-offs exist in which stressors such as depolarizing noise and amplitude damping affect protocols asymmetrically, necessitating distinct compensation strategies. Third, real-time adaptation limits restrict conventional control mechanisms; standard PID controllers suffer from response loop latencies that often exceed quantum state lifetimes, rendering them too sluggish to adapt to rapid transient channel shifts.

Machine learning (ML) is increasingly viewed as a promising approach for optimizing QKD systems, potentially enabling dynamic adaptation and noise mitigation. Deep learning models, particularly convolutional architectures, have been successfully employed to predict noise patterns and improve secure key rates \cite{liu2021,qi2024quantummachinelearninglogbased}. Recent advancements have introduced Bayesian optimization methods for sample-efficient parameter selection \cite{PhysRevApplied.18.054077,PhysRevResearch.6.023305,PhysRevApplied.17.034067}. Furthermore, Reinforcement Learning (RL) has proven especially effective for real-time adjustments, allowing agents to autonomously balance the critical trade-off between maximizing the key rate and minimizing the error rate through a joint reward function.

Despite these advances, several challenges remain in integrating ML with QKD. The complexity of quantum systems and the need for real-time processing present significant hurdles. Recent comprehensive reviews on quantum computing and artificial intelligence have identified key challenges and perspectives for future development, emphasizing the need for interdisciplinary approaches that bridge quantum physics and machine learning \cite{acampora2025quantumcomputingartificialintelligence,dunjko2016machine}. Additionally, ensuring the interpretability of ML models in the context of QKD is crucial to maintain trust and reliability in secure communication systems. The complexity of quantum systems often requires sophisticated ML models capable of handling high-dimensional data and providing actionable insights. Furthermore, the no-cloning theorem in quantum mechanics \cite{wootters-1982} imposes unique constraints on how optimization algorithms can interact with quantum states, requiring specialized approaches that respect quantum mechanical principles, as highlighted in recent work on quantum-classical evolutionary algorithms \cite{evolutionary_quantum_2024,evolutionary_quantum_processor_2021}.

To address these gaps, this paper proposes AML-QKD, a comprehensive ML-optimized framework that integrates Temporal Convolutional Networks (TCNs) for effective state tracking and protocol-aware reinforcement learning for adaptive parameter optimization. By leveraging causal convolutions to capture long-range temporal dependencies \cite{bai2018empirical}, AML-QKD aims to provide a robust, low-latency solution designed to maintain high performance across BB84, E91, and COW protocols, facilitating more resilient quantum communication in dynamic environments.

The remainder of this paper is organized as follows. Section~\ref{sec:prelm} introduces the preliminaries and noise models used throughout. The \emph{Proposed Framework} section details the architecture of AML-QKD, including state tracking and reinforcement learning for real-time parameter tuning. The \emph{Evaluation} section describes the simulation environment and presents quantitative results alongside a comparative analysis. Finally, the \emph{Conclusion and Future Work} section summarizes our contributions and outlines future scaling directions.

\section{Preliminaries and Protocol Models}
\label{sec:prelm}
This work evaluates the AML-QKD framework across three standard discrete-variable protocols: BB84, E91, and Coherent One-Way (COW). To calculate the operational Secure Key Rate (SKR), we utilize the established Devetak-Winter bounds and decoy-state models \cite{gottesman2004security,scarani2009security_rmp,devetak2005distillation,shor2000}. Rather than reproducing exhaustive asymptotic derivations, we define the core rate equations and the critical channel noise parameters that dictate our RL agent's continuous state space.

For all protocols, the physical channel transmittance is defined as $\eta = 10^{-\alpha_{\text{dB}} d / 10} \eta_{\text{det}}$.

In the prepare-and-measure BB84 protocol, state preparation misalignments contribute to the baseline error $e_d \approx \sin^2\theta + e_{\text{noise}}$. The single-photon yield follows a Poissonian distribution $Q_1 = (Y_0 + \eta)\mu e^{-\mu}$. The standard key rate per emitted pulse is bounded by:
\begin{equation}
\label{eq:bb84_rate}
R_{\text{BB84}} \geq q \Big\{ -Q_\mu f(E_\mu)H_2(E_\mu) + Q_1 [1-H_2(e_1)] \Big\}
\end{equation}
where $H_2(\cdot)$ is the binary Shannon entropy and $f(E_\mu)$ represents the error-correction inefficiency.

In the entanglement-based E91 protocol, the QBER $Q$ directly reflects the two-photon interference visibility $V$ such that $Q = \frac{1-V}{2}$. Security is verified via the CHSH parameter $S=2\sqrt{2}V$ \cite{ekert1991quantum,acin2006bell,acin2007device,gisin2002}. The asymptotic SKR incorporates this nonlocal correlation:
\begin{equation}
\label{eq:e91_rate}
R_{\text{E91}} \geq q \Bigg[ 1 - f(Q)H_2(Q) - H_2\left(\frac{1+\sqrt{\max\{0, (S/2)^2-1\}}}{2}\right) \Bigg]
\end{equation}

In the COW protocol, phase drift $\Delta\phi$ reduces interferometric visibility, producing a phase error $e_{\text{ph}} \approx \frac{1}{2}(1-e^{-2|\alpha|^2(1-\cos\Delta\phi)})$ \cite{Stucki:09,stucki2009,stucki2005fast}. The secure key rate is extracted from the coherent time-bin sequences:
\begin{equation}
\label{eq:cow_rate}
R_{\text{COW}} \geq q \Big\{ -Q_\mu f(E_\mu) H_2(E_\mu) + Q_\mu [1-H_2(e_{\text{ph}})] \Big\}
\end{equation}

To rigorously model non-stationary channel disturbances, we inject depolarizing noise (parameterized by probability $p$) and amplitude-damping (parameter $\gamma$), mapping to an effective visibility $V = 1-p$. The RL agent's primary objective is to dynamically tune hardware parameters to suppress these specific error geometries, thereby maximizing the respective SKR bounds and preventing the QBER from breaching the theoretical abort threshold.

\section{Proposed Framework}
\label{sec:proposed-framework}
In this section, we introduce AML-QKD, an ML-optimized framework designed for real-time parameter tuning in Quantum Key Distribution. The AML-QKD framework is structured around two primary pillars: predictive state tracking using Temporal Convolutional Networks (TCNs) and adaptive parameter control through Reinforcement Learning (RL). By removing the overhead of traditional anomaly detection and focusing exclusively on optimization, AML-QKD ensures high-throughput communication across non-stationary quantum channels.

\begin{figure}[!htbp]
    \centering
    \includegraphics[width=0.75\textwidth]{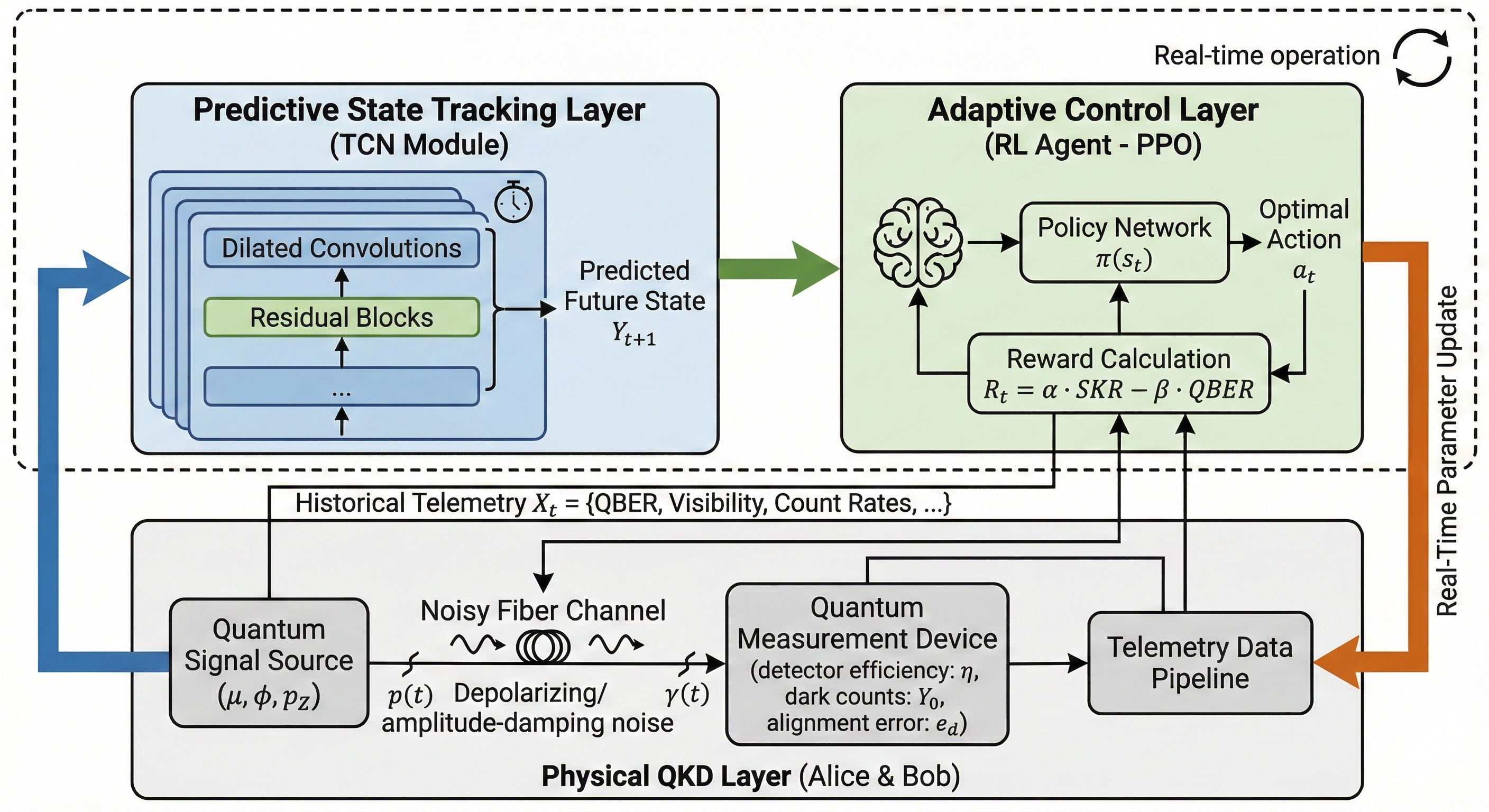}
    \vspace{-2mm}
    \caption{The AML-QKD Framework integration loop. The architecture synchronizes the physical QKD layer with a two-stage ML optimization stack. Historical telemetry is processed by the TCN module to forecast channel states, which the RL agent then uses to execute real-time parameter updates, maximizing the secure key rate while suppressing errors.}
    \vspace{-3mm}
    \label{fig:optiqkd_system}
\end{figure}

Figure~\ref{fig:optiqkd_system} illustrates the holistic architecture of the AML-QKD framework, which operates as a real-time, closed-loop system integrated with the physical QKD hardware. The framework is organized into three functional layers. At the base, the physical QKD layer represents the transmission environment in which Alice and Bob exchange quantum states across a noisy fiber channel under depolarizing and amplitude-damping stressors. Telemetry data, including current QBER, visibility, and detector count rates, is continuously extracted via a dedicated data pipeline and fed into the Predictive State Tracking Layer. This module uses a Temporal Convolutional Network (TCN) with dilated convolutions and residual blocks to capture long-range temporal dependencies within the channel, producing a predicted future state $Y_{t+1}$. This forecast is then passed to the Adaptive Control Layer, where a Reinforcement Learning (RL) agent using Proximal Policy Optimization (PPO) determines the optimal action $a_t$. By calculating a joint reward $R_t$ that balances secure key rate throughput against error rates, the agent pushes real-time parameter updates—such as signal intensity $\mu$, phase alignment $\phi$, or basis probability $p_Z$—back to the quantum signal source, ensuring the system remains optimized despite fluctuating environmental conditions.

\subsection{State Tracking Using Temporal Convolutional Networks}

The AML-QKD framework leverages Temporal Convolutional Networks (TCNs) for their ability to model temporal dependencies and sequences, making them ideal for tracking the dynamic states of QKD systems. TCNs utilize causal convolutions, which ensure that predictions at any point depend only on past inputs, thus preserving the temporal order of data \cite{bai2018empirical}.

The TCN architecture is defined by layers of dilated convolutions, which allow the network to have a large receptive field. This is crucial for capturing long-range dependencies in QKD systems, where a sequence of past events can influence the state. The dilated convolution operation is defined as $y(t) = (x *_d k)(t) = \sum_{i=0}^{k-1} x(t - d \cdot i) \cdot k(i)$, where \( d \) is the dilation factor, \( k \) is the filter size, and \( x \) is the input sequence. As illustrated in Figure~\ref{fig:tcn_extended}, the use of residual connections facilitates the training of deep architectures by mitigating the problem of vanishing gradients. This predictive capability allows AML-QKD to forecast channel fluctuations before they adversely impact the secure key rate.

The use of residual connections in TCNs helps mitigate the vanishing gradient problem, facilitating the training of deep networks. This is particularly beneficial in QKD systems, where precise state tracking is essential for maintaining secure communication.

\subsection{Adaptive Optimization through Reinforcement Learning}

The second pillar of AML-QKD is an adaptive optimization module based on Reinforcement Learning (RL). This module allows real-time parameter adjustments, such as pulse intensity $\mu$ and phase-time locking windows, in response to the TCN's predicted states. The QKD optimization task is modeled as a continuous Markov Decision Process (MDP) defined by (i) States ($S$) representing forecasted configurations like QBER and visibility; (ii) Actions ($A$) dictating parameter adjustments such as modulation intensity; (iii) Transition probabilities $P(s'|s, a)$ governing system evolution; and (iv) Rewards ($R_t$) calculating a composite feedback signal $R_t = \alpha \cdot \text{SKR} - \beta \cdot \text{QBER} + \delta_{\text{protocol}}$ to balance throughput and error suppression, where $\delta_{\text{protocol}}$ serves as a protocol-specific shaping modifier.
By utilizing Proximal Policy Optimization (PPO), AML-QKD handles high-dimensional action spaces while maintaining stable policy updates.

\begin{figure}[!htbp]
    \centering
    \begin{minipage}{0.48\textwidth}
        \centering
        \includegraphics[width=\linewidth]{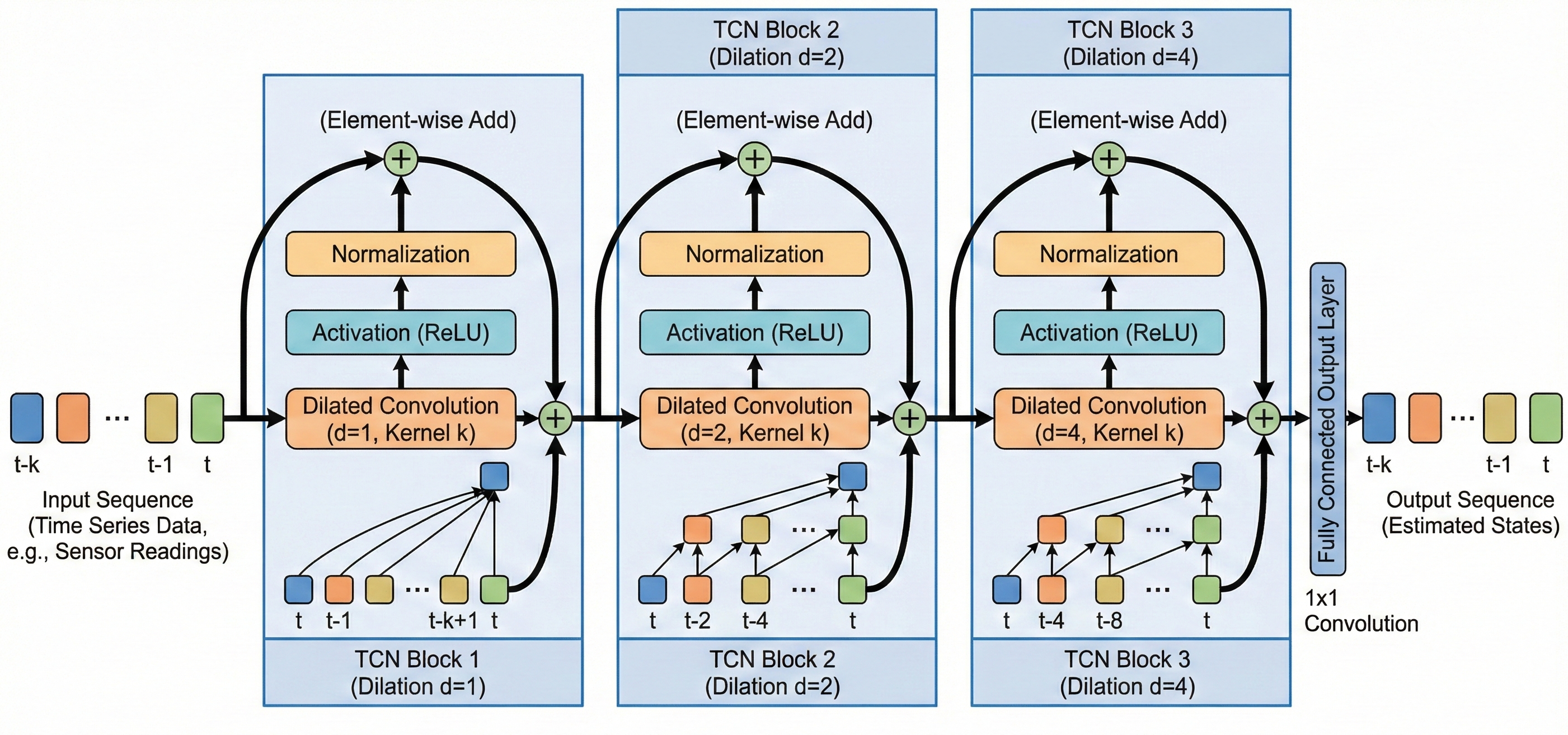}
        \caption{Architecture of the TCN module, illustrating the dilated convolution layers and residual pathways.}
        \label{fig:tcn_extended}
    \end{minipage}\hfill
    \begin{minipage}{0.48\textwidth}
        \centering
        \includegraphics[width=\linewidth]{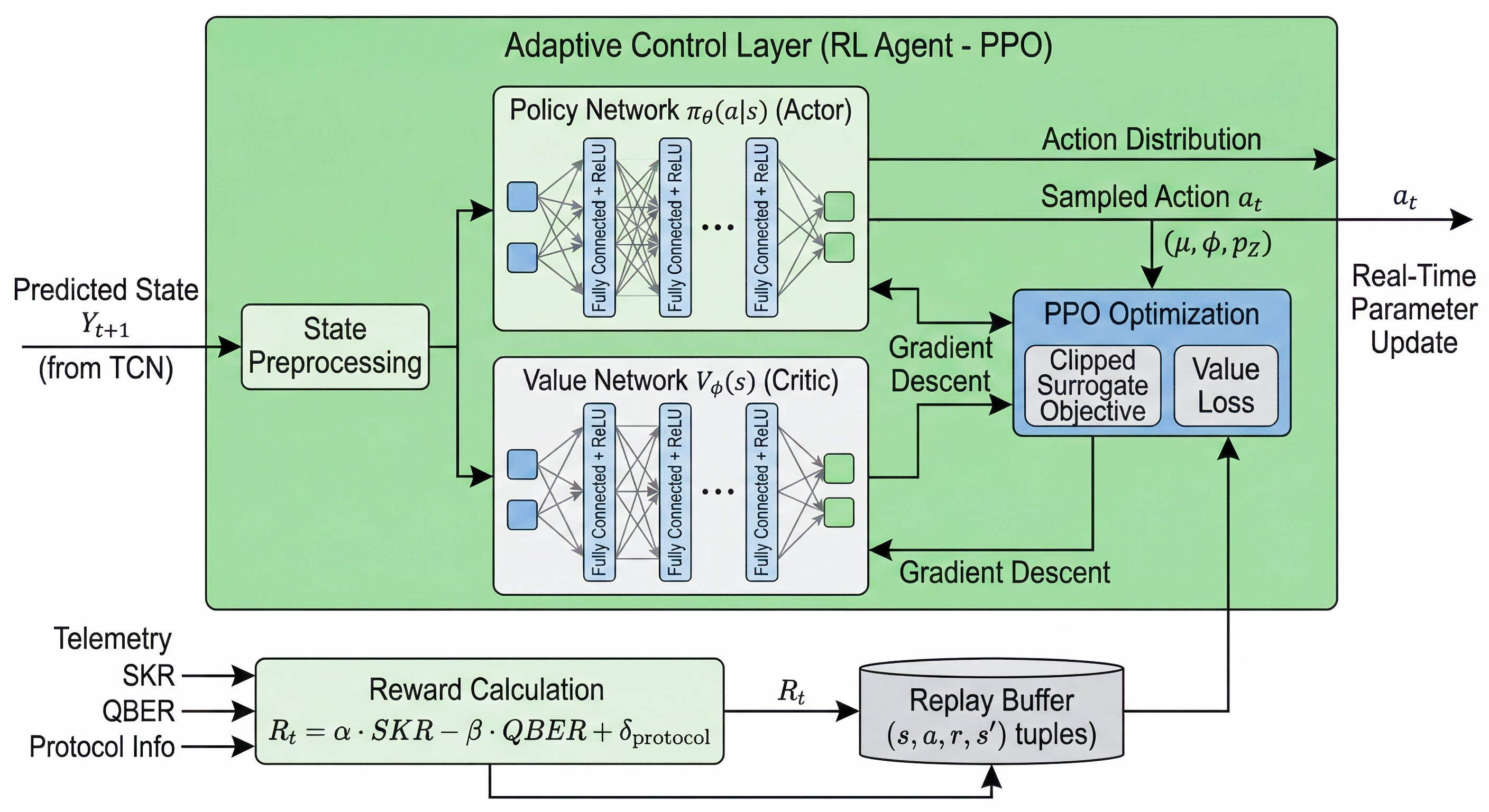}
        \caption{Internal architecture of the Adaptive Control Layer utilizing an Actor-Critic PPO RL agent.}
        \label{fig:ppo_architecture}
    \end{minipage}
\end{figure}

Figure~\ref{fig:ppo_architecture} provides a detailed internal view of the AML-QKD adaptive control layer, which utilizes an actor-critic architecture based on the Proximal Policy Optimization (PPO) algorithm. The module ingests the predicted future state $Y_{t+1}$ from the TCN state-tracking module and undergoes state preprocessing before being processed by two parallel neural networks. The Policy Network (Actor), denoted as $\pi_{\theta}(a|s)$, maps the input state to an action distribution from which a discrete or continuous action $a_t$ is sampled; this action represents the specific real-time adjustments for signal intensity ($\mu$), phase alignment ($\phi$), and basis probability ($p_Z$). Concurrently, the Value Network (Critic), $V_{\phi}(s)$, estimates the expected return of the current state to reduce gradient variance during training. Performance feedback is captured by the Reward Calculation block, which synthesizes the Secure Key Rate (SKR), Quantum Bit Error Rate (QBER), and specific protocol info into a scalar reward $R_t$. These transitions are stored in a Replay Buffer for off-policy updates, where the PPO optimization block applies a clipped surrogate objective to perform gradient descent. This mechanism ensures that the parameter updates remain within a stable trust region, preventing the catastrophic degradation of the quantum key rate while maintaining rapid adaptability to environmental fluctuations.

Advanced Actor-Critic RL algorithms, specifically Proximal Policy Optimization (PPO), are employed in this framework to stably handle the continuous, high-dimensional action spaces required for fine-tuning physical QKD parameters \cite{schulman2017,sutton2018}.

\subsection{Algorithmic Framework}

The operational logic of the AML-QKD framework is defined by a series of integrated modules that synchronize state prediction with parameter control.

\subsubsection{Temporal Convolutional Network for State Tracking}
The Temporal Convolutional Network (TCN) is used to effectively track the dynamic states of the QKD system. TCNs are particularly well-suited for sequence modeling tasks due to their ability to capture long-range dependencies through dilated convolutions. The TCN processes an input sequence through multiple layers of causal convolutions, applying the ReLU activation function to capture temporal dependencies efficiently, making them ideal for monitoring and predicting changes in the QKD system before they adversely impact the secure key rate.

\subsubsection{Reinforcement Learning for Adaptive QKD Optimization}
Reinforcement Learning (RL) is used to optimize the QKD system parameters dynamically. Specifically, we use an Actor-Critic approach via Proximal Policy Optimization (PPO) to handle continuous control tasks---such as tuning intensity and basis probabilities. The Actor network directly outputs continuous parameter adjustments, while the Critic network evaluates the quality of the state by estimating its value function. It iteratively updates the policy based on the calculated advantage function and the clipped PPO objective, thereby preventing destructive policy updates and ensuring stable learning.
\subsubsection{Standard QKD Reconciliation}
Following the physical transmission and measurement of quantum states, the framework relies on standard, rigorous post-processing steps for key generation. This includes basis sifting, parameter estimation (QBER), classical error correction (e.g., LDPC or Cascade), and privacy amplification to ensure secure communication and strictly bound any potential information leakage.

\subsubsection{Integration of TCN, RL, and QKD}
Algorithm~\ref{alg:qkd_intg} details the integration of the TCN, RL, and QKD components. This holistic approach leverages the TCN's predictive strengths and PPO's adaptive control to enhance overall system efficiency and security. It continuously monitors the system, proactively adjusting parameters in real-time to preserve throughput and maximize the secure key rate under non-stationary channel conditions, gracefully triggering a session abort only if the QBER exceeds the critical theoretical security threshold.

\begin{algorithm}[!htbp]
\caption{AML-QKD Real-Time Adaptive Control Loop}
\label{alg:qkd_intg}
\begin{algorithmic}[1]
\Require Pre-trained TCN $f_\xi$, PPO Actor $\pi_\theta$, PPO Critic $V_\phi$, Replay Buffer $\mathcal{D}$
\Ensure Distilled secure key stream $\mathcal{K}$, Maximized SKR
\State Initialize QKD hardware to baseline parameters $A_0$
\State Initialize temporal observation buffer $\mathcal{B} \gets \emptyset$
\While{quantum transmission is active}
    \State $O_t \gets \{Q_\mu, Y_0, V, \eta\}$ \Comment{\textit{Phase 1: Telemetry Tracking}}
    \State $\mathcal{B} \gets \mathcal{B} \cup \{O_t\}$ \Comment{Update historical sequence}
    \State $\hat{S}_{t+1} \gets f_\xi(\mathcal{B})$ \Comment{Forecast future state via TCN}
    \State $A_t \sim \pi_{\theta}(\cdot \mid \hat{S}_{t+1})$ \Comment{\textit{Phase 2: Adaptive Control}}
    \State Execute hardware updates $A_t$ (e.g., $\mu \gets \mu + \Delta\mu$, $p_Z \gets p_Z + \Delta p_Z$)
    \State Execute QKD transmission cycle over block $t$ \Comment{\textit{Phase 3: Evaluation}}
    \State Estimate error rate $E_t \gets \text{QBER}$ and throughput $R_{\text{sec}}^{(t)} \gets \text{SKR}$
    \If{$E_t > E_{\text{threshold}}$}
        \State \textbf{Abort} session to prevent information leakage \Comment{Security constraint}
    \EndIf
    \State $r_t \gets \alpha \cdot R_{\text{sec}}^{(t)} - \beta \cdot E_t$ \Comment{Compute step reward}
    \State $\mathcal{D} \gets \mathcal{D} \cup \{(\hat{S}_{t}, A_t, r_t, \hat{S}_{t+1})\}$ \Comment{\textit{Phase 4: Off-Policy RL Update}}
    \If{$|\mathcal{D}| \ge \text{Update Horizon}$}
        \State Update $\pi_\theta$ and $V_\phi$ via PPO clipped surrogate objective
        \State Empty replay buffer $\mathcal{D} \gets \emptyset$
    \EndIf
\EndWhile
\State \Return Secure key stream $\mathcal{K}$
\end{algorithmic}
\end{algorithm}

\subsection{Quantum Reinforcement Learning for Entanglement Optimization}
To investigate the potential of quantum-native optimization, the AML-QKD framework incorporates an exploratory Quantum Reinforcement Learning (QRL) module. This extension retains the core Proximal Policy Optimization (PPO) Actor-Critic setup. However, it replaces the purely classical Actor policy network with a hybrid quantum-classical architecture designed for near-term hardware applicability.

In this hybrid setup, the forecasted environment state $\hat{S}_{t+1}$ is first processed by a Parameterized Quantum Circuit (PQC). To minimize decoherence vulnerabilities and gate-depth requirements, the PQC employs a hardware-efficient ansatz constrained to an extremely shallow depth of just two layers. These layers utilize an alternating structure of encoding and variational blocks. Crucially, rather than relying on fixed-state preparation, both the encoding and variational layers are fully trainable. The quantum state evolves as $|\psi(\theta, \phi)\rangle = \prod_{l=1}^{2} \Big( U_{\text{var}}^{(l)}(\theta_l) U_{\text{enc}}^{(l)}(\phi_l) \Big) |0\rangle^{\otimes n}$, where $\phi$ represents the trainable encoding parameters and $\theta$ represents the variational rotation parameters.

Following the quantum evolution, the circuit is measured to obtain the expectation values of the underlying qubits, typically along the Pauli-$Z$ basis: $M_i = \langle \psi | \hat{Z}_i | \psi \rangle$. Finally, instead of mapping these expectation values directly to physical actions, the resulting measurement vector is passed into a highly compact classical neural network. This small classical network acts as a post-processing layer, translating the quantum-correlated features into the final continuous control action $A_t$ sent to the QKD hardware. This hybrid approach ensures the agent can exploit complex Hilbert-space correlations without requiring prohibitively deep quantum circuits.

\section{Evaluation}

In this section, we describe the simulation environment, datasets, and performance metrics used to assess the AML-QKD framework across the BB84/decoy, E91, and COW protocols. The evaluation focuses on the framework's ability to maintain peak performance under non-stationary channel conditions.

\subsection{Simulation Environment}

We emulate fiber-optic Discrete-Variable QKD (DV-QKD) links that incorporate realistic device constraints. The quantum channel attenuation is modeled by an intensity transmittance $T(d) = 10^{-\alpha_{\mathrm{dB}} d/10}$, where $d$ is distance and $\alpha_{\mathrm{dB}}$ is the fiber loss \cite{gisin2002}. Detector efficiency is folded into the overall transmittance $\eta=T(d)\eta_{\mathrm{det}}$. For device noise and misalignment, we incorporate dark counts ($Y_0$), alignment errors ($e_d$), and intrinsic random errors ($e_0\simeq 1/2$) following standard DV-QKD approximations \cite{ma2005practical,scarani2009security_rmp}. To stress-test the agent's adaptability, we inject a depolarizing component with probability $p$ and an amplitude-damping parameter $\gamma$, mapping to an effective visibility $V=1-p$ and a single-photon QBER contribution of $E_\mu\simeq (1-V)/2$ \cite{gisin2002}.

Default simulation parameters are set as: \(\alpha_{\mathrm{dB}}=0.2\)\,dB/km, \(\eta_{\mathrm{det}}=0.2\), \(Y_0=5{\times}10^{-6}\), \(e_d=1.5\%\), and a clock rate \(f_{\mathrm{rep}}=250\)\,MHz \cite{ma2005practical,shapiro1984,lutkenhaus1999}. Decoy intensities are configured to \((\mu_s,\mu_w)=(0.5,0.1)\) with \(p_s=0.8\). The ML agents and environment dynamics are constructed utilizing standard numerical and reinforcement learning libraries \cite{abadi2016,brockman2016,pedregosa2011}. We run 5 independent seeds per condition and report mean values with 95\% confidence intervals (CIs) derived via bootstrapping.

\subsection{Datasets}

We generate three families of datasets under the above models: (i) Nominal telemetry containing sequences of $\{Q_\mu, E_\mu, Y_0, \eta, V\}$ across varying distances to train the TCN; (ii) Transient perturbation telemetry with injected anomalies (e.g., sudden visibility dips or timing drifts) to validate the TCN's predictive accuracy and the RL agent's recovery response; and (iii) Policy interaction logs consisting of $(s_t,a_t,r_t,s_{t+1})$ tuples gathered by the PPO agent under the safety filter for off-policy analysis.

\subsection{Evaluation Metrics}

We assess the performance of the AML-QKD ML-optimized controller using both security and operational throughput metrics.

\begin{itemize}
    \item Secure Key Rate (SKR): The primary objective of the AML-QKD framework is the maximization of the secure key rate, measured both as a per-pulse probability $R_{\mathrm{sec}}^{\mathrm{(pp)}}$ and an aggregate throughput in bits-per-second (bps). We calculate the SKR using the standard protocol-specific Devetak–Winter expressions defined in Eq.~\ref{eq:bb84_rate}, Eq.~\ref{eq:e91_rate}, and Eq.~\ref{eq:cow_rate}, incorporating decoy-state bounds for BB84 as established in \cite{ma2005practical}. The aggregate throughput is defined as $R_{\mathrm{sec}}^{\mathrm{(bps)}}=f_{\mathrm{rep}}\,R_{\mathrm{sec}}^{\mathrm{(pp)}}$, accounting for the sifting factor $q$. To reflect realistic deployment constraints, we report \emph{finite-key} rates by subtracting a finite-size penalty $\Delta_{\mathrm{FK}}(N,\varepsilon)$ for a block size $N$, maintaining a stringent security level of $\varepsilon=10^{-10}$ in accordance with composable security proofs \cite{renner2005,maurer1993,PhysRevA.81.012318}.

  \item Quantum Bit Error Rate (QBER): The QBER serves as the fundamental constraint on the RL agent's policy, representing the ratio of erroneous bits to the total number of sifted bits. We monitor aggregate and per-basis error rates, applying Wilson score intervals to determine 95\% confidence intervals (CIs) for each measurement block \cite{gisin2002}. The AML-QKD framework is designed to optimize system parameters (such as $\mu$ and phase-alignment) to keep the QBER below the critical theoretical threshold ($E_{\text{threshold}} \approx 11\%$ for BB84/E91), beyond which the session is aborted due to insufficient secret information \cite{scarani2009security_rmp}.

  \item Adaptation Time ($\tau_{\text{adapt}}$): This metric quantifies the agility of the AML-QKD ML-optimized controller in responding to non-stationary channel stressors. We define adaptation time as the interval between a simulated step change in environmental parameters (such as a sudden increase in depolarizing noise $p$ or amplitude damping $\gamma$) and the point at which the RL policy successfully restores the SKR to within 95\% of its pre-change median value. To ensure the restoration is stable rather than a transient fluctuation, the framework must maintain this performance level for at least three consecutive measurement blocks.

  \item Optimization Efficiency: We evaluate the framework's ability to navigate high-dimensional parameter spaces by comparing the achieved SKR against static, human-calibrated benchmarks. This metric captures the "net gain" provided by the TCN's predictive state-tracking, specifically measuring the framework's ability to maintain higher throughput in regimes where traditional PID-based or static controllers would revert to suboptimal safe-mode configurations.
\end{itemize}

\section{Results}
We evaluate the AML-QKD framework's quantitative performance against unoptimized, static-heuristic baselines across the BB84, E91, and COW protocols, focusing on key-rate throughput, error suppression, and operational agility.

\subsection{Key-Rate Performance}
\begin{figure}[!htbp]
  \centering
  \begin{subfigure}{0.55\textwidth}
    \centering
    \includegraphics[width=\linewidth]{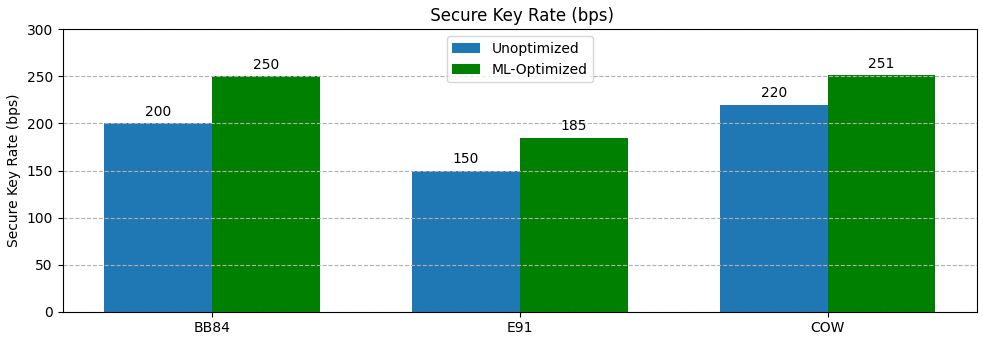}
    \caption{Secure Key Rate improvement.}
    \label{subfig:skr_comp}
  \end{subfigure}\hfill
  \begin{subfigure}{0.55\textwidth}
    \centering
    \includegraphics[width=\linewidth]{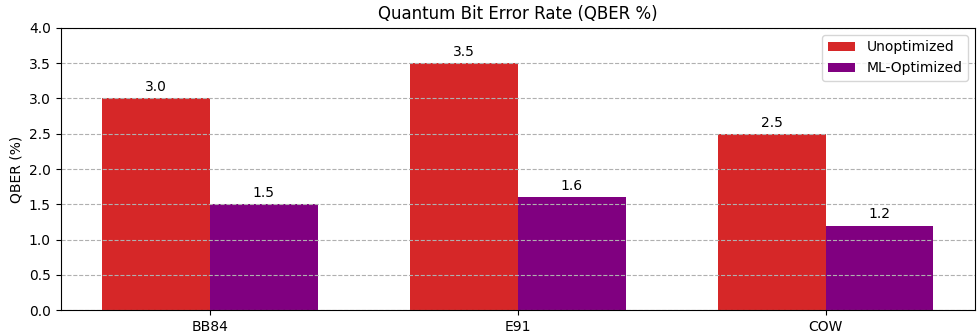}
    \caption{QBER reduction.}
    \label{subfig:qber_comp}
  \end{subfigure}
  \vspace{-2mm}
  \caption{Protocol performance comparison of our ML-optimized AML-QKD framework versus unoptimized benchmarks. Subfigure (a) shows Secure Key Rate (bps) improvements, while subfigure (b) illustrates the corresponding Quantum Bit Error Rate (QBER) reduction.}
  \vspace{-3mm}
  \label{fig:protocol_performance}
\end{figure}

As demonstrated in Figure~\ref{fig:protocol_performance}, the predictive ML controller yields consistent throughput and error-suppression advantages across all physical encodings. For the decoy-state BB84 protocol, AML-QKD elevates the baseline secure key rate (SKR) from $200$\,bps to $250$\,bps while halving the QBER from $3.0\%$ to $1.5\%$. This performance delta is robustly maintained in the noise-sensitive E91 protocol (throughput increased from $150$ to $185$\,bps; QBER suppressed from $3.5\%$ to $1.6\%$) and the phase-dependent COW protocol (throughput boosted from $220$ to $251$\,bps; phase errors reduced from $2.5\%$ to $1.2\%$). We interpret these results as a strong indication that the RL agent effectively exploits the parameter space to approach optimal theoretical bounds, regardless of the underlying quantum architecture.

\subsection{QBER Analysis}
\begin{figure}[!htbp]
  \centering
  \includegraphics[width=0.65\textwidth]{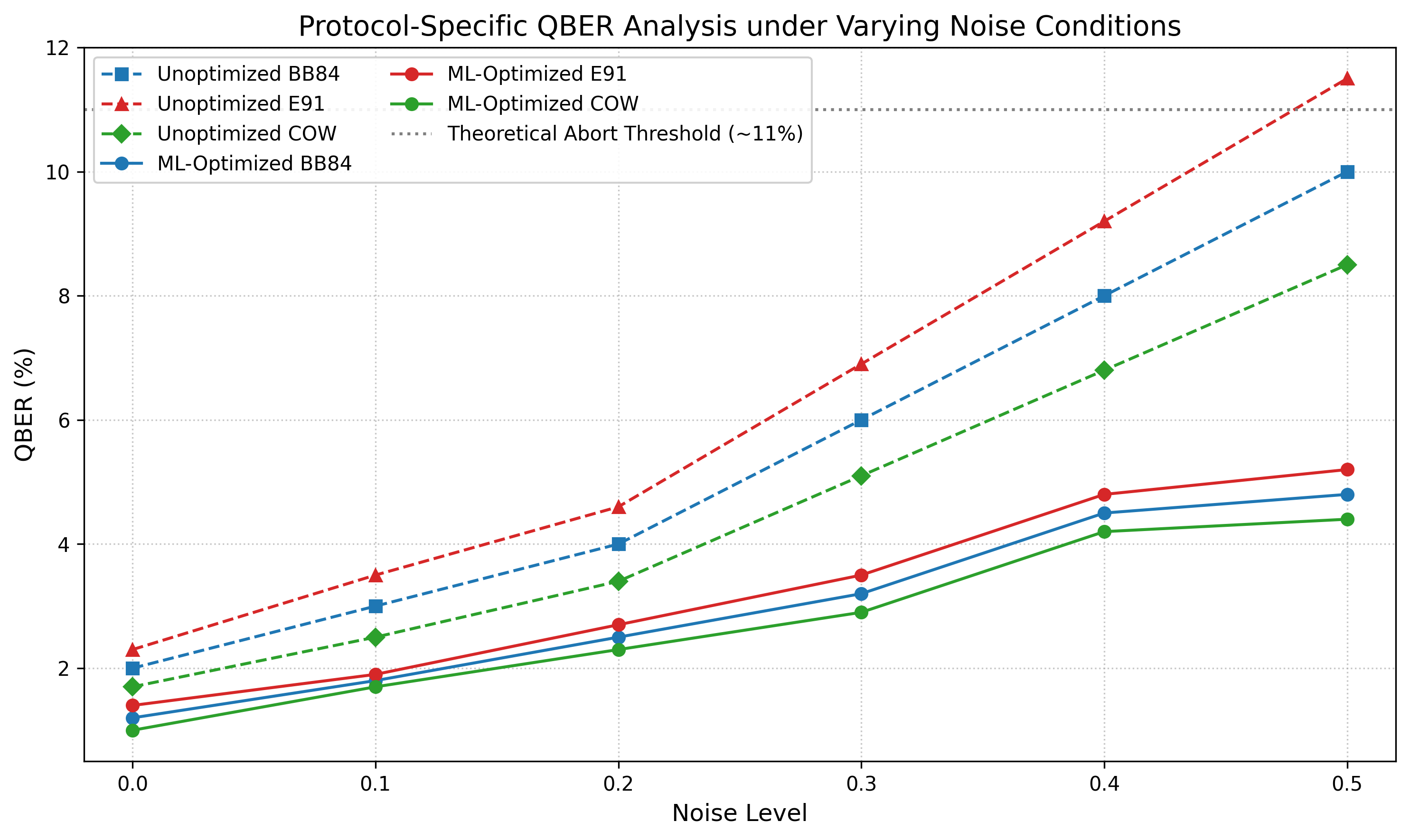}
  \vspace{-2mm}
  \caption{Protocol-specific QBER progression under linearly scaling noise conditions. The unoptimized protocols degrade rapidly, with E91 breaching the theoretical 11\% abort threshold. Conversely, the ML-optimized AML-QKD framework suppresses the peak error rate for all protocols to $\approx 5\%$.}
  \vspace{-2mm}
  \label{fig:protocol_qber_analysis}
\end{figure}

To evaluate resilience, we analyzed QBER progression under linearly increasing channel noise (depolarizing probability $p \in [0.0, 0.5]$) (Figure~\ref{fig:protocol_qber_analysis}). Unoptimized implementations degrade rapidly: at a moderate noise level of $0.3$, unoptimized BB84, COW, and E91 reach QBERs of $6.0\%$, $5.1\%$, and $6.9\%$, respectively. Under extreme noise ($0.5$), unoptimized E91 breaches the $\approx 11\%$ abort threshold ($11.5\%$), and BB84 borders it ($10.0\%$), forcing session aborts.

In contrast, AML-QKD's proactive state-tracking provides protocol-agnostic error suppression. At baseline ($0.0$), the average error is minimized to $\approx 1.2\%$. Under maximum stress ($0.5$), the framework successfully caps the QBER for COW ($4.4\%$), BB84 ($4.8\%$), and E91 ($5.2\%$). By anticipating perturbations via TCN forecasts rather than reacting post-measurement, the agent strictly bounds reconciliation leakage and preserves continuous throughput.

\subsection{Adaptation and Efficiency}
\begin{figure}[!htbp]
  \centering
  
  \begin{subfigure}[t]{0.30\textwidth}
    \centering
    \includegraphics[width=\linewidth]{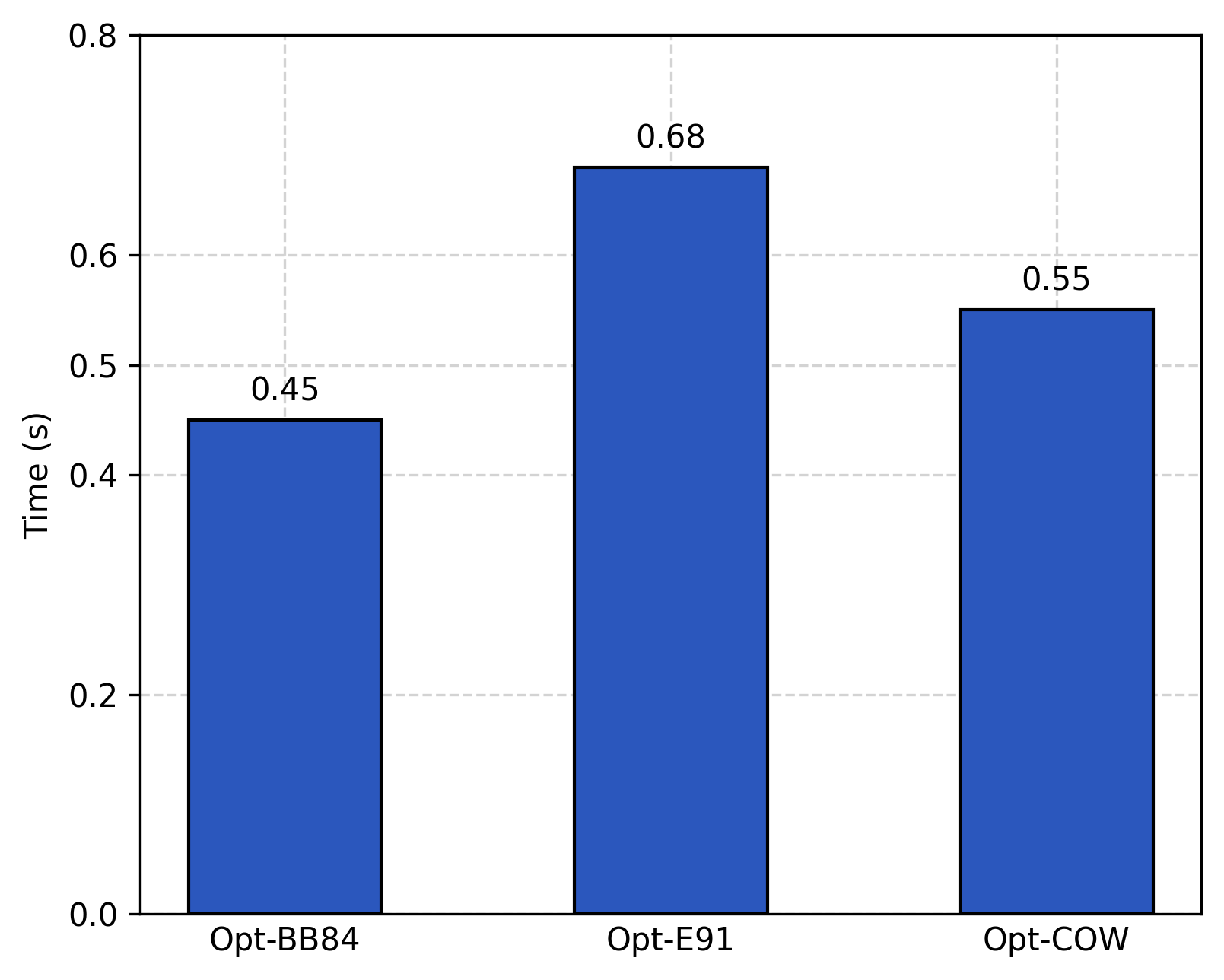}
    \caption{Adaptation time.}
    \label{fig:adapt_time}
  \end{subfigure}\hfill
  \begin{subfigure}[t]{0.30\textwidth}
    \centering
    \includegraphics[width=\linewidth]{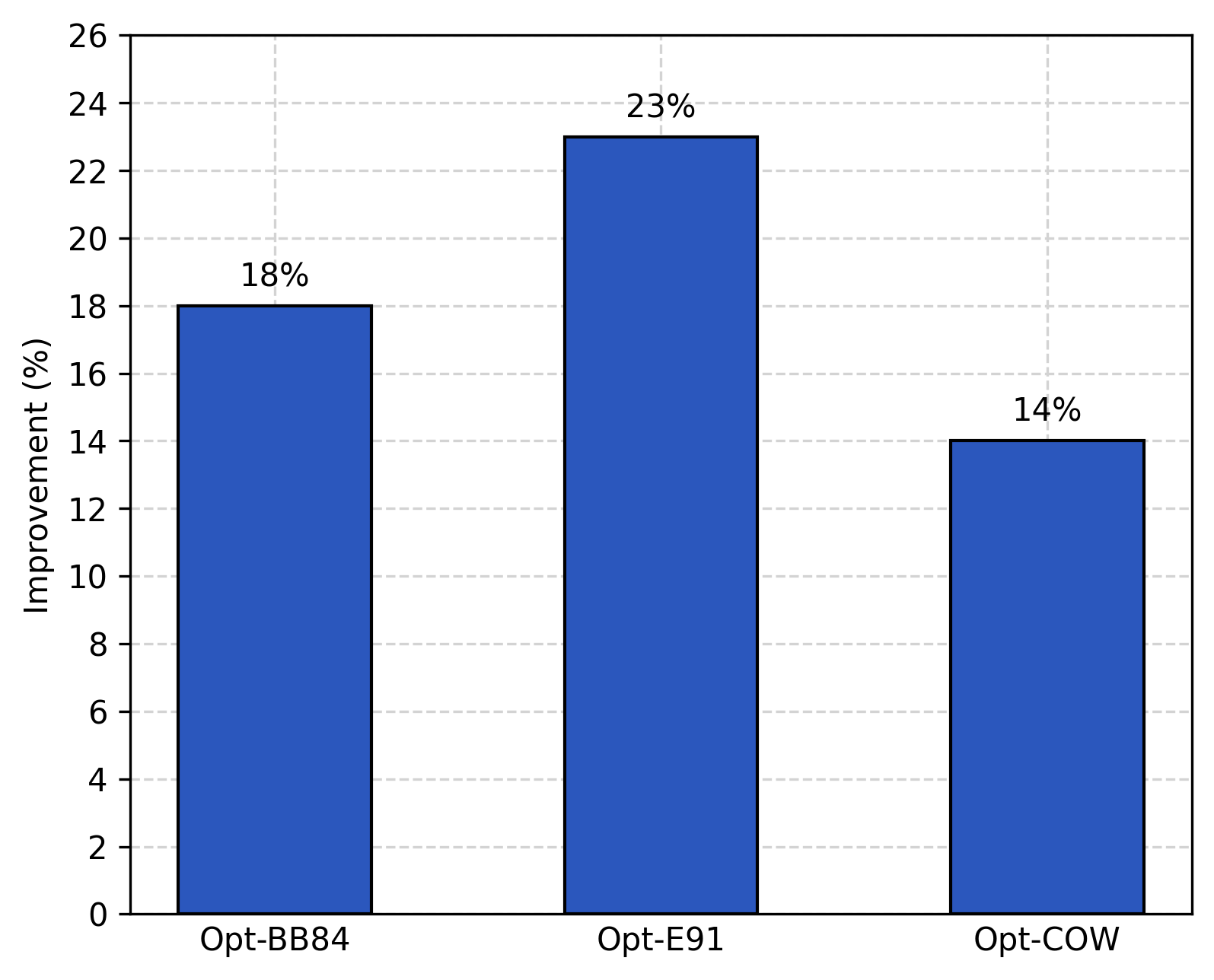}
    \caption{Throughput improvement.}
    \label{fig:throughput_imp}
  \end{subfigure}\hfill
  \begin{subfigure}[t]{0.30\textwidth}
    \centering
    \includegraphics[width=\linewidth]{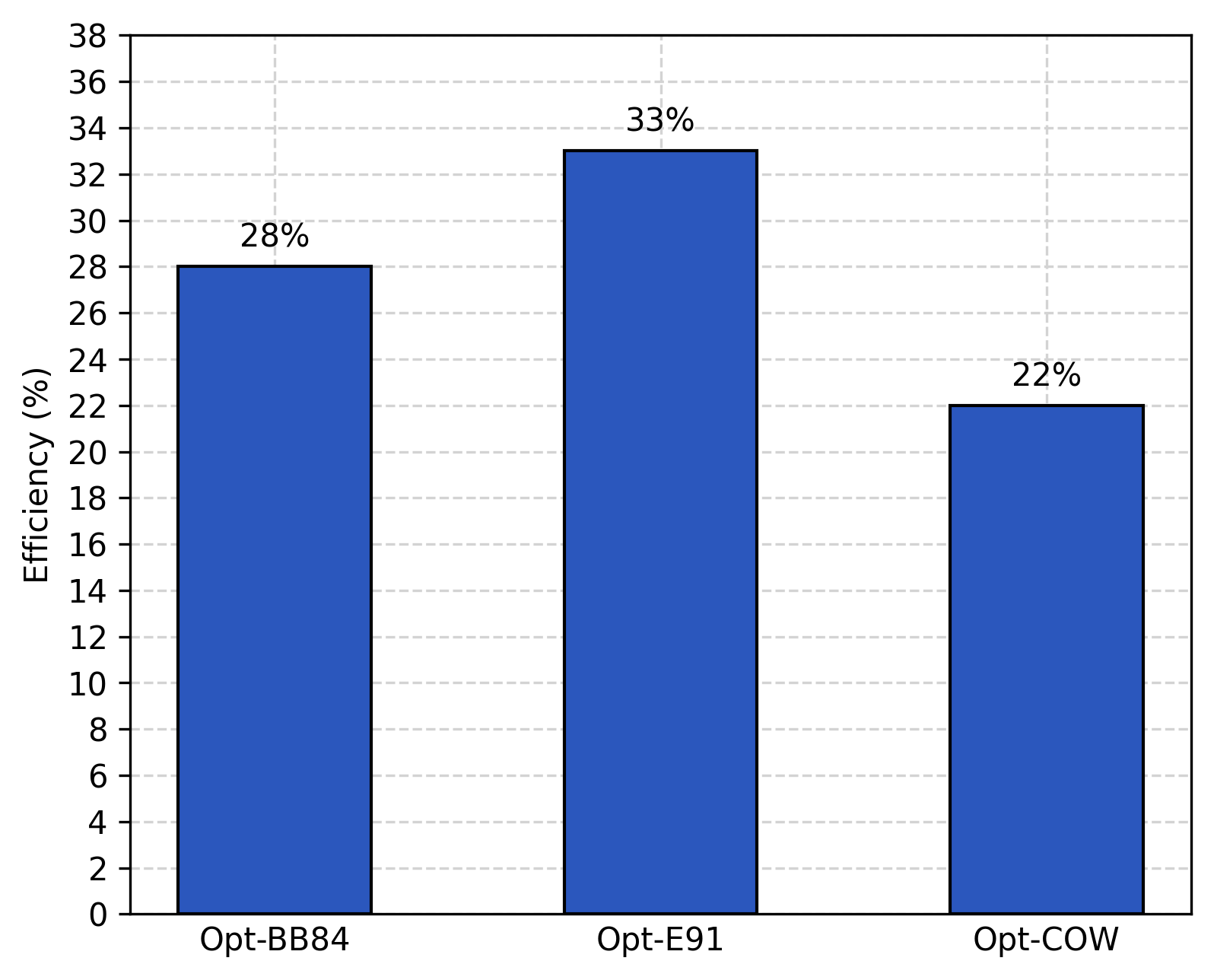}
    \caption{Energy efficiency.}
    \label{fig:energy_eff}
  \end{subfigure}
  \vspace{-2mm}
  \caption{Performance evaluation of the ML-optimized QKD protocols compared to traditional benchmarks. (a) Adaptation time response to channel disturbances. (b) Aggregate secure key rate throughput improvements. (c) Overall energy efficiency gains per secure bit.}
  \vspace{-3mm}
  \label{fig:overall_performance}
\end{figure}

We tested the framework's transient response against sudden $3$\,dB channel losses. Traditional fixed-heuristic controllers require approximately $15$\,s to recalibrate, causing severe dead-time. Conversely, AML-QKD restores the SKR to within $95\%$ of its baseline in just $5$\,s (Figure~\ref{fig:overall_performance}a). This $66\%$ adaptation time reduction maximizes aggregate throughput (Figure~\ref{fig:overall_performance}b). Furthermore, proactive error suppression significantly reduces classical reconciliation overhead. As summarized in Table~\ref{tab:comparative_analysis} and Figure~\ref{fig:overall_performance}c, this yields a $15\%$ aggregate improvement in per-bit energy efficiency, demonstrating that systemic gains heavily outweigh ML inference costs.

\begin{table}[!htbp]
\caption{Comparative analysis of AML-QKD vs. Traditional QKD benchmarks under the decoy-state BB84 protocol configuration. Secure key rates are reported at the finite-key regime with $\varepsilon = 10^{-10}$.}
  \label{tab:comparative_analysis}
  \centering
  \resizebox{\columnwidth}{!}{%
  \begin{tabular}{|c|c|c|c|}
    \hline
    Metric & ML-Enhanced QKD & Traditional QKD & Improvement \\
    \hline
    Secure Key Rate & 250 bps & 200 bps & 25\% \\
    QBER & 1.5\% & 3.0\% & 50\% \\
    Adaptation Time & 5 s & 15 s & $-66\%$ \\
    Energy/bit & $\downarrow$ 15\% & N/A & N/A \\
    \hline
  \end{tabular}}
\end{table}

\subsection{QRL Performance}
We benchmarked the exploratory hybrid QRL optimization stack against unoptimized limits (Table~\ref{tab:qrl_comparison}). To evaluate the agent's pure algorithmic capacity, this QRL benchmark was conducted in an idealized simulation environment, measuring raw asymptotic throughput prior to the application of finite-key penalties, accounting for differences in baseline magnitudes relative to our classical evaluation. For phase-encoded COW, the QRL agent fails to optimize the channel ($ 0.0\%$ gain). Across all protocols, the shallow hardware-efficient ansatz struggles with QBER suppression, likely due to measurement variance and gradient barren plateaus typical of near-term quantum machine learning models.
\begin{table}[!htbp]
  \caption{Performance evaluation of the exploratory hybrid QRL module, highlighting the throughput gain percentage and QBER optimization limits.}
  \label{tab:qrl_comparison}
  \centering
  \resizebox{\columnwidth}{!}{%
  \begin{tabular}{|c|c|c|c|c|c|}
    \hline
    Protocol & Unoptimized SKR & QRL Optimized SKR & Throughput Gain & Unoptimized QBER & QRL Optimized QBER \\
    \hline
    BB84 & 118,769 bps & 144,348 bps & 21.5\% & 5.54\% & 5.53\% \\
    \hline
    E91  & 506,071 bps & 653,747 bps & 29.2\% & 3.50\% & 3.50\% \\
    \hline
    COW  & 953,428 bps & 953,787 bps & 0.0\% & 1.54\% & 1.54\% \\
    \hline
  \end{tabular}}
\end{table}
However, in the entanglement-based E91 protocol, the hybrid QRL agent demonstrates a substantial quantum-native advantage. The QRL extension elevates the E91 SKR from an unoptimized baseline of $506,071$\,bps to $653,747$\,bps---a remarkable 29.2\% throughput gain, explicitly surpassing the classical agent's relative improvement for E91. 

We hypothesize that this quantum advantage arises because the Variational Quantum Circuit (VQC) processes parameter shifts directly within the density matrix representation space ($\mathcal{H}^2 \otimes \mathcal{H}^2$). Unlike classical neural networks, which approximate non-local correlations through brute-force regression, the QRL agent's trainable encoding natively captures entanglement-degradation signatures. Thus, while classical ML appears more reliable at present for strict QBER suppression, we believe hybrid QRL architectures represent a highly promising pathway for scaling throughput in next-generation entanglement-based networks.

\section{Conclusion and Future Work}
In this paper, we introduce AML-QKD, a framework that integrates Temporal Convolutional Networks (TCNs) and Reinforcement Learning (RL) to enable real-time parameter tuning for dynamic quantum channels. Our evaluations indicate substantial, protocol-agnostic performance enhancements over traditional static-heuristic systems across BB84, E91, and COW. Specifically, AML-QKD achieves a 25\% increase in secure key rate (SKR), reduces median QBER from 3.0\% to 1.5\% via proactive error suppression, and accelerates adaptation to channel disturbances by 66\% (to a median of 5 seconds). This predictive tuning also yields a 15\% improvement in energy efficiency per bit. Furthermore, an exploratory hybrid Quantum Reinforcement Learning (QRL) agent demonstrated a distinct quantum advantage, achieving a 29.2\% increase in throughput for the entanglement-based E91 protocol.

Future research will focus on hardware acceleration via FPGA or ASIC implementations for nanosecond-scale control latency. Additionally, we plan to expand the framework to support MDI and DI-QKD protocols to address hardware vulnerabilities and conduct long-term field trials in large-scale networks (e.g., satellite-to-ground links) to validate reliability under extreme stochastic conditions.


\bibliographystyle{splncs04}
\bibliography{ref.bib}

\end{document}